\documentclass[preprint,aps,showkeys,preprintnumbers,onecolumn,showpacs,amsmath,amssymb,floatfix,endfloats*]{revtex4}
\usepackage{graphicx}
\usepackage{dcolumn}
\usepackage{bm}
\begin{document}
\preprint{\textit{Version July 30th 07}}
\title{Highly efficient single photon emission from single quantum dots within a two-dimensional photonic bandgap}
\author{M. Kaniber}
\email{kaniber@wsi.tum.de}
\author{A. Laucht}
\author{T. H\"urlimann}
\author{M. Bichler}
\author{R. Meyer}
\author{M.-C. Amann}
\author{J. J. Finley}
\affiliation{Walter Schottky Institut, Technische Universit\"at M\"unchen, Am Coulombwall 3, D-85748 Garching, Germany}
\date{\today}
\begin{abstract}
We report highly efficient single photon generation from InGaAs self-assembled quantum dots emitting within a two-dimensional photonic bandgap. A strongly suppressed multiphoton probability is obtained for single quantum dots in bulk GaAs and those emitting into the photonic bandgap. In the latter case, photoluminescence saturation spectroscopy is employed to measure a $\sim$17 times enhancement of the average photon extraction efficiency, when compared to quantum dots in bulk GaAs. For quantum dots in the photonic crystal we measure directly an external quantum efficiency up to 26\%, much higher than for quantum dots on the same sample without a tailored photonic environment. The results show that highly efficient quantum dot single photon sources can be realized, without the need for complex nanopositioning techniques.
\end{abstract}
\maketitle
%
%
Practical and highly efficient single photon sources are a fundamental prerequisite for elementary quantum optics, quantum cryptography \cite{Gisin02} and linear optical quantum computation \cite{Knill01}. Such devices should emit one and only one photon on demand at a defined frequency and with high external quantum efficiency ($\eta$). Over the last decade single photon sources have been demonstrated using many different approaches, such as single atoms \cite{Kuhn02}, molecules \cite{Brunel99, Lounis00}, color centers in solids \cite{Beveratos02}, and semiconductor quantum dots (QDs) \cite{Santori01}. However, most of these approaches suffer from a very low $\eta$ limiting the potential advantages they offer, when compared to attenuated coherent pulses. The highest values of $\eta$ have been achieved using QDs coupled to various kinds of optical microresonators, such as microdisks \cite{Michler00}, microposts \cite{Pelton02}, and photonic crystal (PC) nanocavities \cite {Chang06}. However, the coupling of single QDs to nanocavity modes is technologically challenging, requiring precise spatial and spectral tuning of the QD emission. Here, we propose a much simpler route towards high $\eta$ single photon sources. Our approach is based on photonic bandgap (PBG) materials (without cavities) to realize efficient single photon sources that can be used for applications in quantum optics and quantum information processing.

%
%
In this Letter we present detailed optical studies of single self-assembled In$_{0.5}$Ga$_{0.5}$As QDs, both inside and outside a photonic environment created by a two-dimensional (2D) PC nanostructure. QDs inside the PC are shown to emit photons much more efficiently when compared to those in the unpatterned substrate. This effect is shown to be due to the efficient spatial redistribution of the spontaneous emission (SE) caused by the 2D-PBG \cite{Yablonovitch87}. Photon correlation measurements performed on QDs in bulk GaAs and in PCs both exhibit clear photon antibunching, proving the single photon character of the emission. Power dependent photoluminescence (PL) measurements recorded with pulsed excitation reveal that photons emitted from QDs in the PC can be collected up to $\sim$~17 times more efficiently than those from QDs in bulk GaAs. Furthermore, we measure a $\eta\sim$~26\% which demonstrates the great potential of single QDs in PCs as highly efficient and practical single photon sources for quantum optics experiments and quantum information processing.\\
%
%
Our structure consists of an undoped GaAs substrate onto which a 500 nm thick Al$_{0.8}$Ga$_{0.2}$As sacrificial layer was deposited by molecular beam epitaxy. Following this, a 180 nm thick GaAs waveguide was grown with a single layer of self-assembled InGaAs QDs embedded at the midpoint. PC nanostructures consisting of a triangular lattice of air holes were subsequently fabricated using standard e-beam lithography and reactive ion etching. In a final process step a free-standing GaAs membrane was established by HF wet chemical etching. Full details of the sample structure and processing techniques can be found in Ref. \cite{Kaniber07}.\\
%
%
The sample was mounted in a liquid He-flow cryostat (15 K) and excited by 2 ps duration optical pulses delivered from a mode-locked Ti-sapphire laser at a repetition rate of $f_{laser}=$~80 MHz. The excitation wavelength was chosen to be $\lambda_{exc}=$~850 nm, into the wetting layer beneath the QDs. The emission from the sample was collected by a 100 $\times$ microscope objective (NA=0.8) providing submicron spatial resolution and was analyzed using a 0.5 m imaging monochromator. For detection, we used a Si based charged coupled device (CCD) camera for $\mu$-PL experiments, a single silicon avalanche photo diode (temporal resolution of $\sim$~350 ps) for time-resolved spectroscopy or a pair of similar detectors in Hanbury Brown and Twiss \cite{Hanbury56} configuration setup for measuring the temporal statistics of the SE from single QDs \cite{Loudon83}.\\ 
%
\begin{figure}[t]
    \begin{center} 
       \includegraphics[width=\columnwidth]{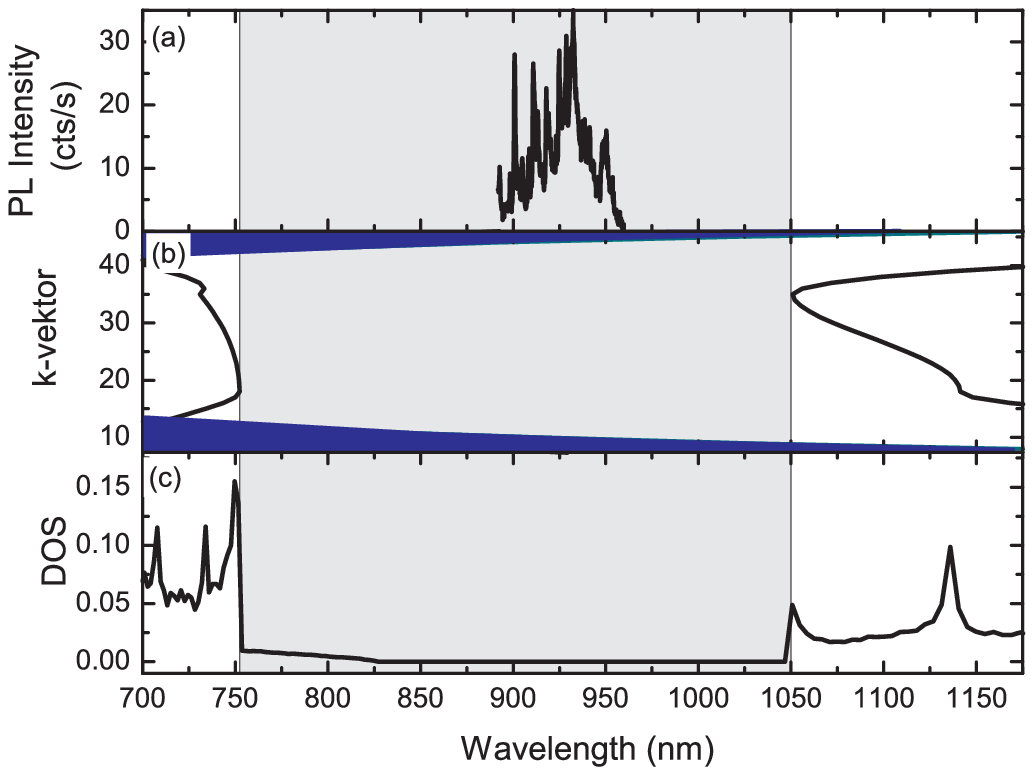}
    \end{center}    
    \caption{ (color online) (a) $\mu$-PL spectrum recorded from a QD ensemble. (b) Three-dimensional calculation of the photonic bandstructure for the GaAs 2D-PC slab with r/a = 0.335, lattice constant a = 280 nm and thickness d = 180 nm. The dark blue region denotes the light line. (c) Corresponding photonic density of states as a function of the wavelength. The grey region denotes the 2D-PBG.}
\end{figure}
%
%
Under weak optical excitation the QD ensemble emits between 890 nm and 960 nm as shown in Fig. 1a. The QDs have properties which are typical for InGaAs QDs, e.g. linear and quadratic behavior for the intensity of single exciton (X) and biexciton (2X) transitions on excitation power, respectively. Typical SE decay lifetimes  for excitons were around 0.6 ns as found previously for ensemble measurements \cite{Kaniber07}. To ensure that the QDs emit into the PBG, we calculated the three-dimensional bandstructure for a GaAs 2D-PC with \textit{r/a}=0.335, where \textit{r} is the radius of the air holes and \textit{a}=280 nm the lattice constant of the PC. The simulated \textit{r/a}-ratio is obtained from scanning electron microscopy images of the investigated PCs. The calculated photonic bandstructure is presented in Fig. 1b, showing the continuum photonic bandedges (black solid lines), the appearance of a 2D-PBG (gray shaded region) for TE-like polarized light from $\sim$750 nm to $\sim$1050 nm, and the light cone (blue shaded region). The emission of the QD ensemble lies spectrally deep inside the PBG, such that pronounced cavity quantum electrodynamic effects are expected \cite{Kaniber07}. To access the strength of these effects we calculated the photonic density of states for the same parameter space and obtained a strong suppression of the total photonic density of states within the spectral region of the PBG (Fig. 1c). The 2D-PBG gives rise to a decreased number of optical states in the plane of the PC into which the QDs can emit. Therefore, the QD emission is spatially redistributed and directed perpendicular to the sample surface, which can be collected more efficiently \cite{Kaniber07}.\\  
%
\begin{figure}[t]
    \begin{center} 
       \includegraphics[width=\columnwidth]{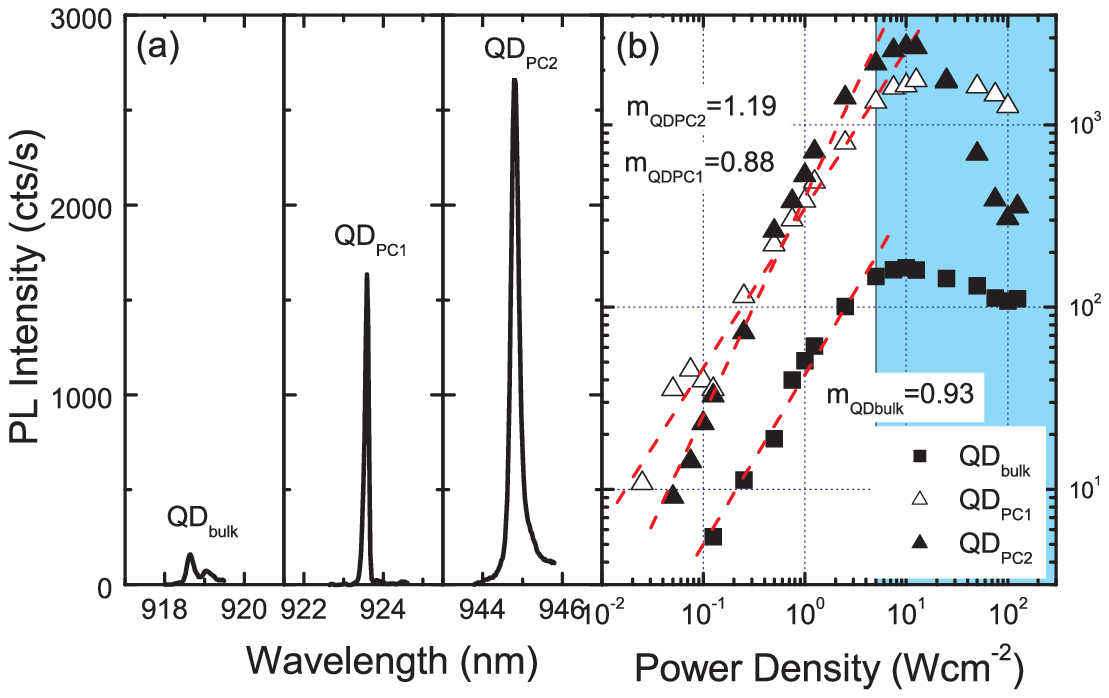}
    \end{center}
    \caption{ (color online) (a) $\mu$-PL spectra of single QDs in bulk GaAs (QD$_{bulk}$) and in a PC nanostructure (QD$_{PC1}$, QD$_{PC2}$) recorded with an excitation power of $10~Wcm^{-2}$. (b) Power dependent measurements for all three QDs indicating ground state emission due to the linear behavior. The blue shaded region indicates the saturation regime.}
\end{figure}
%
%
In Fig. 2a we compare $\mu$-PL measurements for a typical single QD measured next to the PC in bulk GaAs (QD$_{bulk}$, $\lambda_{QD_{bulk}}=918.64$~nm) and a pair of single QDs embedded in the PC (QD$_{PC1}$, $\lambda_{QD_{PC1}}=923.59$~nm; QD$_{PC2}$, $\lambda_{QD3_{PC2}}=944.80$~nm). These data were recorded using intermediate excitation power ($10~Wcm^{-2}$) and a CCD multichannel detector. Clearly, the PL intensities of QD$_{PC1}$ and QD$_{PC2}$ appear to be much higher than QD$_{bulk}$ when measured under similar excitation conditions. This observation already indicates that the 2D-PBG serves to spatially redistribute the SE from QD$_{PC1}$ and QD$_{PC2}$, when compared to QD$_{bulk}$ \cite{Fujita05}. In order to prove this expectation, we studied the saturation behavior of the intensity of the three QDs as a function of the excitation power density per pulse. The results of these measurements are presented in Fig. 2b.\\
At low excitation powers($<5~Wcm^{-2}$) we observe a linear behavior for all three QDs (with slopes $m_{QD_{bulk}}=0.93$, $m_{QD_{PC1}}=0.88$, and $m_{QD_{PC2}}=1.19$), clearly showing that these lines arise from the radiative recombination of a single electron hole pair (exciton) in the QDs. In strong contrast, at higher excitation power we observe a pronounced saturation of the $\mu$-PL intensity. In this regime ($>10~Wcm^{-2}$, light blue region in Fig. 2b), each excitation pulse should give rise to one, and only one, photon at the exciton transition wavelength. Under pulsed excitation the maximum optical power ($P_{QD}^{max}$) generated by a QD is given by:
\begin{equation}
P_{QD}^{max}=\hbar\omega_{QD}\cdot f_{laser}.
\label{eqn1}
\end{equation}
Where $\omega_{QD}$ is the emitted photon frequency and $f_{laser}$ the laser repetition rate. In such a scenario an excitation laser with $f_{laser}=$~80 MHz should result in a maximum photon count rate $\Gamma_{QD}^{max}=P_{QD}^{max}/\hbar\omega_{QD}=8\times10^7$ cps emitted from a single QD. In reality, only a small fraction of the emitted photons are detected in our experiment due to emission into guided modes of the GaAs slab, the combined optical losses in the collection system and finite detector detectivity. This reduction of the maximum photon count rate $\Gamma_{QD}^{max}$ is directly observed in the powerseries (PS) in Fig. 2b by the decreased values of the saturation level $\Gamma_{QD_{bulk}}^{PS}=165\pm8$~cps, $\Gamma_{QD_{PC1}}^{PS}=1637\pm82$~cps, and $\Gamma_{QD_{PC2}}^{PS}=2700\pm135$~cps, for QD$_{bulk}$, QD$_{PC1}$, and QD$_{PC2}$, respectively. Therefore, we deduce a relative extraction enhancement ($\sigma$) for QD$_{PC1}$ and QD$_{PC2}$ in the PC compared to QD$_{bulk}$ in bulk GaAs from the ratios $\sigma_{QD_{PC1}}=\Gamma_{QD_{PC1}}^{PS}/\Gamma_{QD_{bulk}}^{PS}=9.9\pm1.0$ and $\sigma_{QD_{PC2}}=\Gamma_{QD_{PC2}}^{PS}/\Gamma_{QD_{bulk}}^{PS}=16.4\pm1.7$, respectively. This means that we can collect light from a QD in a PC nanostructure up to $\sim$17 times more efficiently than from a QD in bulk GaAs. The difference in the extraction enhancement $\sigma_{QD_{PC1}}$ when compared with $\sigma_{QD_{PC2}}$ arises from a change of the local photonic density of states \cite{Nikolaev07}.\\
The value of $\sigma$ extracted above represents the relative enhancement of the absolute photon extraction efficiency $\eta$ due to the PBG. We now continue to measure the absolute value of $\eta$ for the investigated QDs determining the combined photon count rates ($\Gamma^{Corr}$) on both single photon counters of a HBT setup for QD$_{bulk}$, QD$_{PC1}$, and QD$_{PC2}$. Thus, we obtain $\Gamma_{QD_{bulk}}^{Corr}=3000\pm300$ cps, $\Gamma_{QD_{PC1}}^{Corr}=20000\pm2000$ cps, and $\Gamma_{QD_{PC2}}^{Corr}=40000\pm4000$ cps, respectively, for an excitation power of $10~Wcm^{-2}$, where all QDs are close to the saturation regime (c.f. Fig. 2b). Quite generally, $\eta$ is obtained by dividing the total photon count rate $\Gamma_{QD}^{Corr}$ on both detectors of the HBT by the detection efficiency of the setup $\rho_{detection}$, the quantum efficiency of the detector $\phi_{detector}$ ($\phi_{detector}^{bulk}=26.3\%\pm0.6\%$, $\phi_{detector}^{PC1}=28.2\%\pm0.6\%$, $\phi_{detector}^{bulk}=21.5\%\pm0.4\%$), and the laser repetition rate $f_{laser}$:
\begin{equation}
\eta_{QD}=\frac{\Gamma_{QD}^{Corr}}{f_{laser}\cdot\rho_{detection}\cdot\phi_{detector}}.
\label{eqn2}
\end{equation}
The detection efficiency $\rho_{detection}$ of our measurement setup was carefully measured by sending laser light, tuned to the emission wavelength of the three QDs, into the optical detection system. Using a calibrated optical powermeter, we compared the optical power reaching the avalanche photo diode (APD) detector with the optical power entering the collection objective. This procedure resulted in an absolute detection efficiency $\rho_{detection}=0.875\pm0.1$\%.\\
Inserting this value in eqn. (2) results in $\eta_{QD_{bulk}}=1.6\%\pm0.2\%$, $\eta_{QD_{PC1}}=10.1\%\pm1.0\%$ and $\eta_{QD_{PC2}}=26.6\%\pm~2.7\%$ for QD$_{bulk}$, QD$_{PC1}$ and QD$_{PC2}$, respectively. By comparing these values of $\eta$ with $\eta_{QD_{bulk}}$, we obtain an enhancement of the extraction efficiencies of $6.2\pm1.8\times$ for QD1 and $16.3\pm0.9\times$ for QD2, respectively. These values are in excellent agreement with the values presented above for the relative enhancement $\sigma$ estimated from the power dependent measurements, supporting the validity of our simple analysis.\\
%
\begin{figure}[t]
    \begin{center}
      \includegraphics[width=\columnwidth]{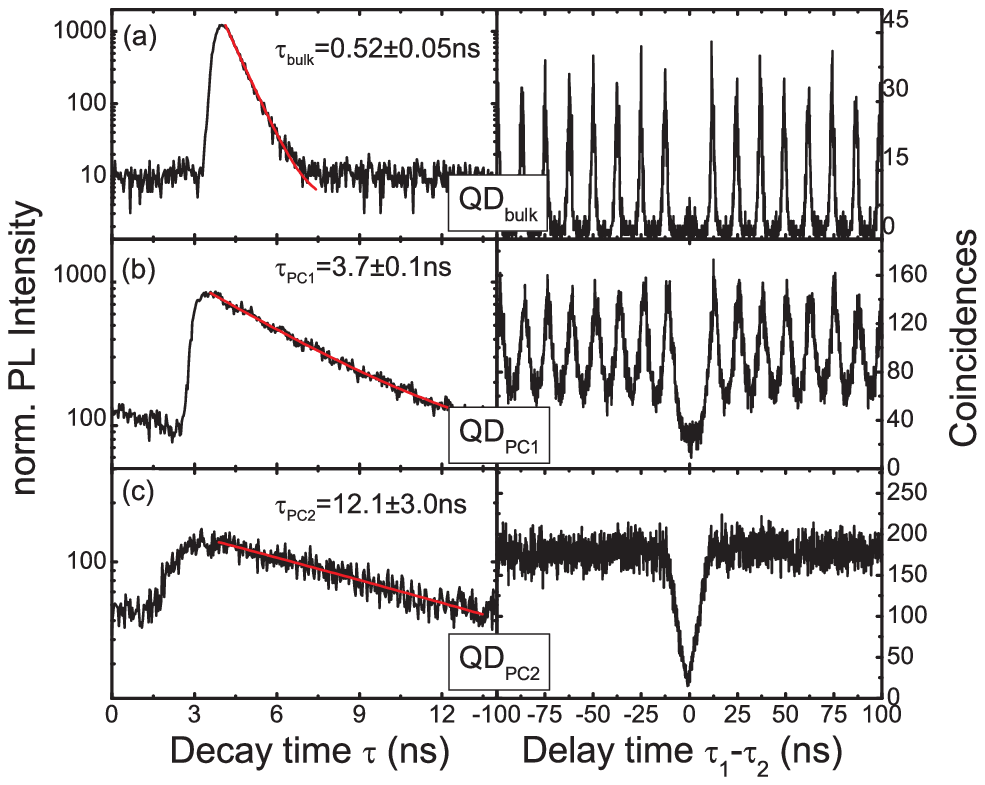}
    \end{center}
    \caption{ (color online) (Left panel) Time-resolved PL data for QD$_{bulk}$ (a), QD$_{PC1}$ (b), and QD$_{PC2}$ (c), respectively. (Right panel) Corresponding photon correlation measurements for QD$_{bulk}$ (a), QD$_{PC1}$ (b), and QD$_{PC2}$ (c), demonstrating clear signature of photon antibunching.}
\end{figure}
%
%
In Fig. 3a (left panel) we present a time-resolved $\mu$-PL spectrum of QD$_{bulk},$ from which we extract an exciton lifetime of $\tau_{bulk}=0.52\pm0.05$ ns, in good accordance with typical values for InGaAs QDs \cite{Michler03}. The single-photon character of the QD emission is demonstrated by measuring the second-order correlation function $g^{(2)}(\tau)$. For the pulsed Ti:sapphire laser we obtain a series of equally spaced peaks which correspond to the laser repetition rate (not shown here). However, for a single photon emitter the peak at zero delay time should vanish \cite{Kimble77}. The measured histogram of QD$_{bulk}$ as a function of the delay time $\tau_1-\tau_2$ between the two detectors (Fig. 3a (right panel)) shows clear signature of photon antibunching from which we determine a reduction of the multiphoton probability to be less than $\sim$19\%.\\
Similarly, we demonstrate the single photon character of QD$_{PC1}$ and QD$_{PC2}$ inside the PC. QD$_{PC1}$ exhibits a significantly longer lifetime $\tau_{PC1}=3.7\pm0.1$ ns (Fig. 3b (left panel)) which originates from the reduced photon density of states inside the PBG \cite{Kress05}. Nevertheless, the correlation measurement (Fig. 3b (right panel)) indicates strong multiphoton suppression due to the absence of the peak at zero delay time. The enhanced background between the adjacent peaks arises due to the lifetime lengthening of the QD transition, which is reflected in the correlation spectrum by the width of the pulses. This effect is even more pronounced for QD$_{PC2}$ which has a lifetime $\tau_{PC2}=12.1\pm3.0$ ns (Fig. 3c (left panel)), which is comparable to the time between two adjacent laser pulses ($1/f_{laser}=12.5 ns$). Therefore, we no longer observe peaks in the correlation measurements (Fig. 3c (right panel)) but a flat line, similar to photon correlation measurements under continuous wave excitation. The continuous wave like characteristic of QD$_{PC2}$ is also observed in its power series (Fig. 2b, black triangles), which shows a decrease of the PL intensity for high pumping powers instead of the saturation expected for pulsed excitation \cite{Finley01}. The antibunching dip at $\tau_1-\tau_2=$~0 ns still proves the single photon nature of the emission. The observation of a higher value of $\eta$ for QD$_{PC2}$ compared with QD$_{PC1}$ combined with the lower SE-rate, indicates that non-radiative processes can be neglected in our experiment.\\
The observation of pronounced antibunching in the photon correlation measurements combined with the measurement of the absolute external quantum efficiencies show that QDs embedded in PC nanostructures are suitable for efficient single photon generation. The enhanced emission is in very good agreement with systematic time-resolved PL experiments performed on InGaAs QD ensembles in PC nanostructures \cite{Kaniber07}. Such an efficient single photon source is a promising candidate to be used in quantum cryptography. When compared to recently demonstrated single photon generation from QDs in microcavities \cite{Hennessy07, Press07}, our approach is technologically less demanding, since we do not rely on deterministic positioning \cite{Badolato05} and spectral tuning of emitter \cite{Kiraz01} or mode \cite{Strauf06}.\\
%
%
In summary, we have presented efficient single photon generation from QDs inside a PBG material. When compared to QDs in bulk GaAs ($\eta_{QD_{bulk}}=~$1.6\%), the incorporation of QDs into a PC nanostructure enhances the quantum efficiency by a factor up to $\sim 17\times$ and results in an absolute extraction efficiency for QDs in  the PC of $\eta_{QD_{PC2}}=~$26\%. This is to our knowledge the highest reported experimental value for the external quantum efficiency in PC nanostructures. Furthermore, this value can easily be further enhanced, for example by growing a distributed Bragg reflector below the PC slab waveguide in order to reduce the losses of the photons emitted towards the GaAs substrate.\\
We acknowledge financial support of the Deutsche Forschungsgemeindschaft via the Sonderforschungsbereich 631, Teilprojekt B3 and the German Excellence Initiative via the "Nanosystems Initiative Munich (NIM)".\\

\end{document}